\pdfoutput=1

\documentclass[a4paper,conference]{IEEEtran}

\usepackage[noadjust]{cite}      

\usepackage{graphicx}  
\usepackage{comment}

\usepackage{subfig}

\usepackage{float} 

\usepackage{url}       

\usepackage{amsmath}   
\usepackage{fancyheadings}
\begin{document}

\title{Behavior of Oil under Breaking Waves by \\a Two-phase SPH Model}

\author{\IEEEauthorblockN{Zhangping Wei}
\IEEEauthorblockA{Department of Civil Engineering\\
Johns Hopkins University\\
Baltimore, MD 21218, USA\\
zwei@jhu.edu; zwei.coast@gmail.com}
\and
\IEEEauthorblockN{Huabin Shi}
\IEEEauthorblockA{Department of Hydraulic Engineering\\
Tsinghua University\\ Beijing 100084, China\\
shb11@mails.tsinghua.edu.cn}
\and
\IEEEauthorblockN{Cheng Li}
\IEEEauthorblockA{Department of Mechanical Engineering\\
Johns Hopkins University\\
Baltimore, MD 21218, USA\\
chengli2.718@gmail.com}
\and
\IEEEauthorblockN{Joseph Katz}
\IEEEauthorblockA{Department of Mechanical Engineering\\
Johns Hopkins University\\
Baltimore, MD 21218, USA\\
katz@jhu.edu}
\and
\IEEEauthorblockN{Robert A. Dalrymple}
\IEEEauthorblockA{Department of Civil Engineering\\
Johns Hopkins University\\
Baltimore, MD 21218, USA\\
rad@jhu.edu}
\and
\IEEEauthorblockN{Giuseppe Bilotta}
\IEEEauthorblockA{Istituto Nazionale di Geofisica e Vulcanologia\\
Sezione di Catania, 95125 Catania, Italy\\
giuseppe.bilotta@ingv.it}}

\maketitle

\begin{abstract}
A two-­phase Smoothed Particle Hydrodynamics (SPH) model has been developed on the basis of GPUSPH, which is an open­-source implementation of the weakly compressible SPH method on graphics processing units, to investigate oil dispersion under breaking waves. By assuming that the multiple phases are immiscible, the two-phase model solves the same set of governing equations for both phases. Density in each phase is preserved by renormalization, and the harmonic mean of viscosities is used in the transition zone. Interfacial surface tension effect between the oil and the water is considered by a numerical surface tension model. The model is first used to simulate a single oil drop rising through still water. The numerical model predicts realistic shape change of the oil drop during the rising process caused by the buoyancy force. Next it is applied to reproduce a laboratory experiment on oil dispersion under breaking waves conducted at Johns Hopkins University. Several high­-speed cameras were used to record the interaction between breaking waves and the oil. Comparison with the laboratory measurements shows that GPUSPH is able to reproduce well the pre- \& post-breaking wave in the laboratory; however, oil dispersion predicted by GPUSPH only match part of the laboratory observation. Several factors (e.g., 3D \& chaotic nature of breaking waves, numerical setup) cause the discrepancy.
\end{abstract}

\section{Introduction}
Oil released offshore due to an oil spill can be transported nearshore by wind waves among many other processes. In the nearshore zone, the breaking waves facilitate oil break-up and mixing throughout the water column, reaching the seabed. There is no denying about the importance of studying oil dispersion under breaking waves, such as understanding the ecological impact of oil spills, and supporting decision making. There are several challenges when simulating oil dispersion under breaking waves by a numerical model. One is that the model has to be able to consider multi-phase interactions involving interfacial surface tension effect. The other is that the numerical model should be able to robustly simulate the complicated 3D breaking waves. This work addresses this issue by using a Smoothed Particle Hydrodynamics (SPH) model GPUSPH, which has proven effective for simulating breaking waves (see, e.g., \cite{dalrymple2006numerical}, \cite{wei2016numerical}, \cite{wei2015sph}, and \cite{wei2016simulation}), and it has been recently improved to consider multi-phase interaction. 

In terms of multi-phase SPH models, there are several methods available in literature to preserve the multi-phase interface, where the density discontinuity and viscosity discontinuity exist. Hu and Adams \cite{hu2006multi} proposed a multi-phase SPH method from a particle smoothing function in which the neighboring particles only contribute to the specific volume but not density. Their method is able to handle density discontinuity and also capable of considering multi-viscosity, multi-surface tension interaction. After their work, the specific volume approach for multi-phase modeling has been followed by several works (see, e.g., \cite{grenier2009hamiltonian}). However, based on the experience of the authors of this study, it is tedious to implement the specific volume approach together with a moving boundary in a two-phase SPH model. Another straightforward approach is to directly solve the density as an unknown for multiple phases (see, e.g., \cite{adami2012generalized}, \cite{monaghan2005smoothed}, and \cite{shakibaeinia2011mesh}). However, this method is limited to a relatively small density ratio, and special attention is needed when the simulation involves a free surface boundary.

In this study, we analyze the behavior of oil under breaking waves by simulating a laboratory experiment conducted by Li \& Katz~\cite{li2016} at Johns Hopkins University, by using a two-phase SPH model. The rest of the paper is organized as follows. The governing equations of the two-phase SPH method are introduced in Section~\ref{governing_equations}. Section~\ref{numerical_examples} presents two numerical examples on oil-water interaction simulated by the two-phase model. The laboratory experimental setup of oil dispersion under breaking waves and the corresponding numerical setup are introduced in Section~\ref{lab_model}. Then Section~\ref{results_discussions} shows the comparison between the laboratory measurement and the numerical predictions on breaking waves and oil dispersion, and it also discusses the causes of the discrepancy. Finally, conclusions are made in Section~\ref{conclusions}.

\section{Two-phase SPH model}\label{governing_equations}
\subsection{Governing equations}
The SPH method is used to solve the Navier--Stokes equations, which are given by
\begin{equation}
\frac{\mathrm{D} \rho}{\mathrm{D} t} = - \rho \nabla\cdot \mathbf{u}\label{eq:cn}
\end{equation}
\begin{equation}
\frac{\mathrm{D} \mathbf{u}}{\mathrm{D} t} = -\frac{\nabla P}{\rho} + \mathbf{g} + \frac{1}{\rho}\nabla \cdot (\mu \nabla\mathbf{u}) +\frac{\mathbf{F}_s}{\rho} \label{eq:me}
\end{equation}
where $t$ is time; $\rho$ is fluid density; $\mathbf{u}$ is particle velocity; $P$ is pressure; $\mathbf{g}$ is the gravitational acceleration; $\mu$ is the dynamic viscosity; $\mathbf{F}_s$ is the surface tension force per volume.

As the fluid is assumed to be weakly compressible in this study, the pressure in Eq.~(\ref{eq:me}) can be directly computed by using the equation of state \cite{monaghan1992smoothed} by
\begin{equation}
P=\beta \left [ \left ( \frac{\rho}{\rho_{0}} \right )^{\gamma }-1 \right ]\label{eq:state}
\end{equation}
where $\rho_{0}$ is the initial density; $\gamma$ is chosen to be 7; and the parameter $\beta$ is calculated by
\begin{equation}
\beta = \frac{\rho_{0}C_{s0}^{2}}{\gamma} \label{beta}
\end{equation}
where $C_{s0}$ is the speed of sound evaluated with $\rho = \rho_{0}$. The real speed of sound restricts the numerical time step to be very small, resulting in high computation cost. To alleviate this limitation, Monaghan~\cite{monaghan1994simulating} suggested to use a reduced speed of sound, which satisfies $C_{s0} / u_{max} > 10$ (where $u_{max}$ is the maximum velocity in the simulation) to avoid unphysical density fluctuation. 

\subsection{Discretization of the equation of continuity}
Considering a single phase in the domain, the equation of continuity (\ref{eq:cn}) is discretized by
\begin{equation}
\frac{\mathrm{d}\rho_i}{\mathrm{d}t} = \sum_{j=1}^{n}m_j\mathbf{u}_{ij}\cdot \nabla_i W_{ij}\label{eq:ct_dist_1}
\end{equation}
where $i$ is the index of the particle of interest; $j$ is the index of the neighboring particle of particle $i$; $n$ is the total number of neighboring particles of particle $i$; $m$ is the particle mass; the velocity difference vector is defined by $\mathbf{u}_{ij} = \mathbf{u}_{i}-\mathbf{u}_{j}$; and $\nabla_i W_{ij}$ is the gradient of the kernel function $W$ at particle $i$.

When it comes to two-phase flows, the above discretization (\ref{eq:ct_dist_1}) is modified as
\begin{equation}
\frac{\mathrm{d}\rho_i}{\mathrm{d}t} = \rho_i\sum_{j=1}^{n}\frac{m_j}{\rho_j}\mathbf{u}_{ij}\cdot \nabla_i W_{ij}\label{eq:ct_dist_2}
\end{equation}

By comparing with Eq. (\ref{eq:ct_dist_1}),  Eq. (\ref{eq:ct_dist_2}) considers the impact of the volume instead of the mass of neighboring particles on the density variation of particle $i$, and it is able to simulate two-phase flows interactions (e.g., \cite{monaghan2005smoothed,adami2012generalized}). If there are particles with different phases inside the kernel of particle $i$, the actual amount of same phase particles (${n}'$) should be used to compute the density, and then the kernel function contribution/weighting is normalized since ${n}' < n$. One has
\begin{equation}
\rho_i = \frac{\sum_{j=1}^{{n}'}m_j W_{ij} }{ \sum_{j=1}^{{n}'}\frac{m_j}{\rho_j}W_{ij}}\label{eq:ct_dist_3}
\end{equation}

Note that Eq. (\ref{eq:ct_dist_3}) is necessary for simulations involving a free surface, otherwise, Eq. (\ref{eq:ct_dist_2}) is enough for handling a two-phase problem. 
\subsection{Discretization of the equation of motion}
The current two-phase formulation discretizes the pressure term in a symmetric way:
\begin{equation}
-\left ( \frac{\nabla P}{\rho} \right )_i = -\sum_{j}^{n}m_j\left ( \frac{P_i}{\rho_i^2} +\frac{P_j}{\rho_j^2}\right ) \nabla_i W_{ij}\label{eq:dist_p_1}
\end{equation}

The above discretization conserves both linear and angular momentums. 

In the single-phase implementation of GPUSPH, the viscosity term is discretized by following Lo and Shao \cite{lo2002simulation}:
\begin{equation}
\left ( \nabla \cdot (\nu_{0} \nabla\mathbf{u}) \right )_i = \sum_{j}^{n}m_j \frac{4\nu_{0}}{\rho_i + \rho_j} \frac{\mathbf{r}_{ij}\cdot\nabla_iW_{ij}}{ \| \mathbf{r}_{ij}  \|^2+\delta ^2}\mathbf{u}_{ij} \label{eq:kvis}
\end{equation}
where $\nu_{0}$ is the laminar kinematic viscosity; $\delta$ is a small number introduced to keep the denominator non-zero and usually equal to 0.1$h$ ($h$ is the smoothing length).

For multiple phases, the viscous term can be discretized in two ways, depending on the used viscosity types. When the kinematic viscosity ($\nu = \frac{\mu}{\rho}$) is used, one has
\begin{equation}\label{eq:visc_1}
\begin{split}
\left ( \nabla \cdot (\nu \nabla\mathbf{u}) \right )_i & = \sum_{j}^{n}\frac{m_j}{\rho_j}\left (\frac{4\nu_{i}\nu_{j}}{\nu_i + \nu_j}\right )\frac{\mathbf{r}_{ij}\cdot \nabla_iW_{ij}}{\left \| \mathbf{r}_{ij} \right \|^2+\delta ^2} \mathbf{u}_{ij} \\ &=\sum_{j}2\nu_{ij}\frac{m_j}{\rho_j}\frac{\mathbf{r}_{ij}\cdot \nabla_iW_{ij}}{\left \| \mathbf{r}_{ij} \right \|^2+\delta ^2} \mathbf{u}_{ij}
\end{split}
\end{equation}
On the other hand, discretization with the dynamic viscosity reads
\begin{equation}\label{eq:visc_2}
\begin{split}
\left (\frac{1}{\rho} \nabla \cdot (\mu \nabla\mathbf{u}) \right )_i & = \sum_{j}^{n}\frac{m_j}{\rho_i\rho_j}\left (\frac{4\mu_{i}\mu_{j}}{\mu_i + \mu_j}\right )\frac{\mathbf{r}_{ij}\cdot \nabla_iW_{ij}}{\left \| \mathbf{r}_{ij} \right \|^2+\delta ^2} \mathbf{u}_{ij} \\
& =\frac{1}{\rho_i}\sum_{j}2\mu_{ij}\frac{m_j}{\rho_j}\frac{\mathbf{r}_{ij}\cdot \nabla_iW_{ij}}{\left \| \mathbf{r}_{ij} \right \|^2+\delta ^2} \mathbf{u}_{ij}
\end{split}
\end{equation}
where $\nu_{ij}$ and $\mu_{ij}$ are viscosities in the form of the harmonic mean \cite{shakibaeinia2011mesh}:
\begin{equation}
 \nu_{ij}= \frac{2\nu_{i}\nu_{j}}{\nu_i + \nu_j}
\end{equation}
\begin{equation}
 \mu_{ij}= \frac{2\mu_{i}\mu_{j}}{\mu_i + \mu_j}
\end{equation}
For a single phase the harmonic mean viscosity recovers the single phase viscosity, and then Eqs.~(\ref{eq:visc_1}) and (\ref{eq:visc_2}) become the same as Eq.~(\ref{eq:kvis}). However, the harmonic mean viscosity provides a smoother approximation of the two-phase viscosity in the transition zone. In this study, Eq.~(\ref{eq:visc_1}) was used in the numerical simulation.

To model the surface tension, we utilize the continuous surface force model of Brackbill et al.~\cite{brackbill1992continuum}, and the formulation of Hu and Adams~\cite{hu2006multi} is implemented in this study. First, a color-index function $C$ is used to differentiate the multiple fluids, and $C$ of a particle $i$ related to a phase $x$ is defined by:
\begin{equation}
C_{i}^{x}=\left\{
\begin{matrix}
1 & i\in x  \\ 
0 & i\notin x
\end{matrix}\right.\label{eq:C}
\end{equation}
Furthermore, the gradient of the color-index $\nabla C$ has a zero value for particles within the same phase, but it has a unit jump across the interface.

The surface tension force is written in a tensor form as $\mathbf{F}_s=\nabla\cdot\mathbf{T}$, where the surface stress tensor $\mathbf{T}$ is given by
\begin{equation}
\mathbf{T}=\alpha_s\frac{1}{\nabla C}\left ( \frac{1}{d}\mathbf{I} \left \|  \nabla C\right \|^2 -\nabla C  \otimes \nabla C\right )\label{eq:csf}
\end{equation}
where $\alpha_s$ is the interfacial surface tension coefficient; $d$ is the spatial dimension, and $d = 3$ in this study; $\mathbf{I}$ is the identity matrix; and the sign $\otimes$ indicates the outer product of two vectors.

Considering a two-phase scenario with phase indices $x$ and $y$, the color-index gradient of particle $i$ of phase $x$ exists if there is a neighboring particle $j$ of phase $y$:
\begin{equation}
\nabla C_{i}^{xy} = \sigma_i\sum_{j}^{n}\left [ \frac{C_i^y}{\sigma_i^2} +\frac{C_j^y}{\sigma_j^2}\right ]\nabla_iW_{ij} \qquad  x\neq y; i\in x;j\in y \label{eq:cd}
\end{equation}
where $\sigma = \rho / m$. According to the definition of $C$ in Eq. (\ref{eq:C}), $C_i^y \equiv  0$. Next, we can determine Eq. (\ref{eq:csf}) with Eq. (\ref{eq:cd}), and the surface tension force tensor is discretized by
\begin{equation}
\left (  \frac{\mathbf{F}_s}{\rho}\right )_i = \frac{1}{m_i}\sum_{j}^{n}\left ( \frac{T_{a,b}^i}{\sigma_i^2} +\frac{T_{a,b}^j}{\sigma_j^2}\right ) \cdot \nabla_i W_{ij}
\end{equation}

The above governing equations of the two-phase model have been implemented on the basis of a single-phase SPH code GPUSPH (www.gpusph.org.), which is an open­-source implementation of the weakly compressible SPH method on graphics processing units~\cite{herault2010sph}. Recent efforts have been made to extend GPUSPH to run on multiple GPUs~\cite{rustico2012smoothed}, and achieve uniform precision~\cite{herault_modelling_2014}. As a results, it could be used to simulate large-scale processes with a high resolution (see, e.g., recent applications of GPUSPH in the field of coastal engineering in \cite{wei2016numerical}, \cite{wei2015sph} and \cite{wei2016simulation}).

\section{Numerical examples on oil-water interaction}\label{numerical_examples}

\begin{figure}
\centering
\includegraphics[width=2.75in]{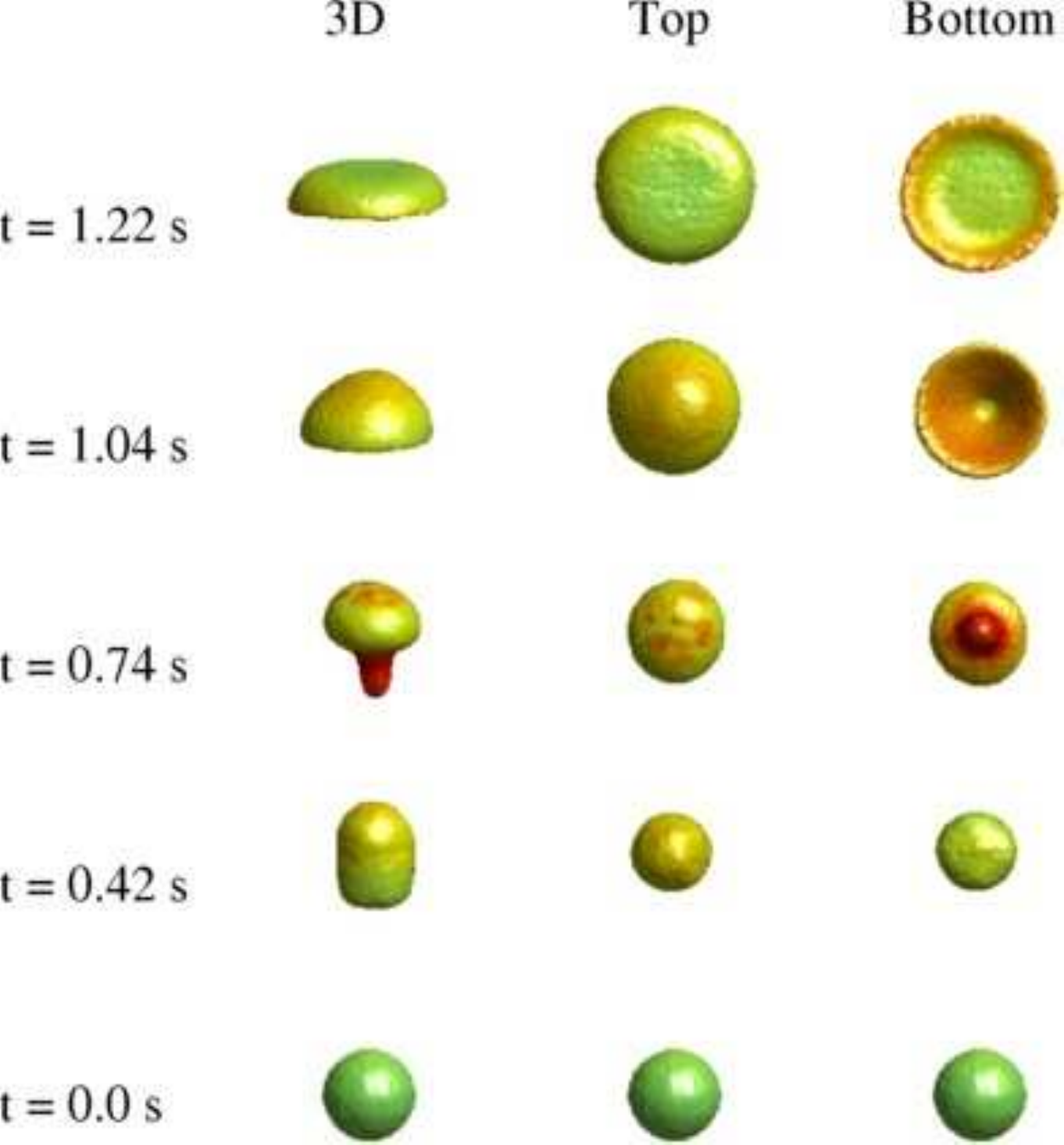}
\caption{GPUSPH simulation of an initially spherical oil drop with diameter $d$ = 0.04~m rising through still water with a water depth of 0.2~m. 3D view (left column); top view (middle column); bottom view (right column).}
\label{fig:oil_sphere}
\end{figure}

\begin{figure}
\centering
\includegraphics[width=2.75in]{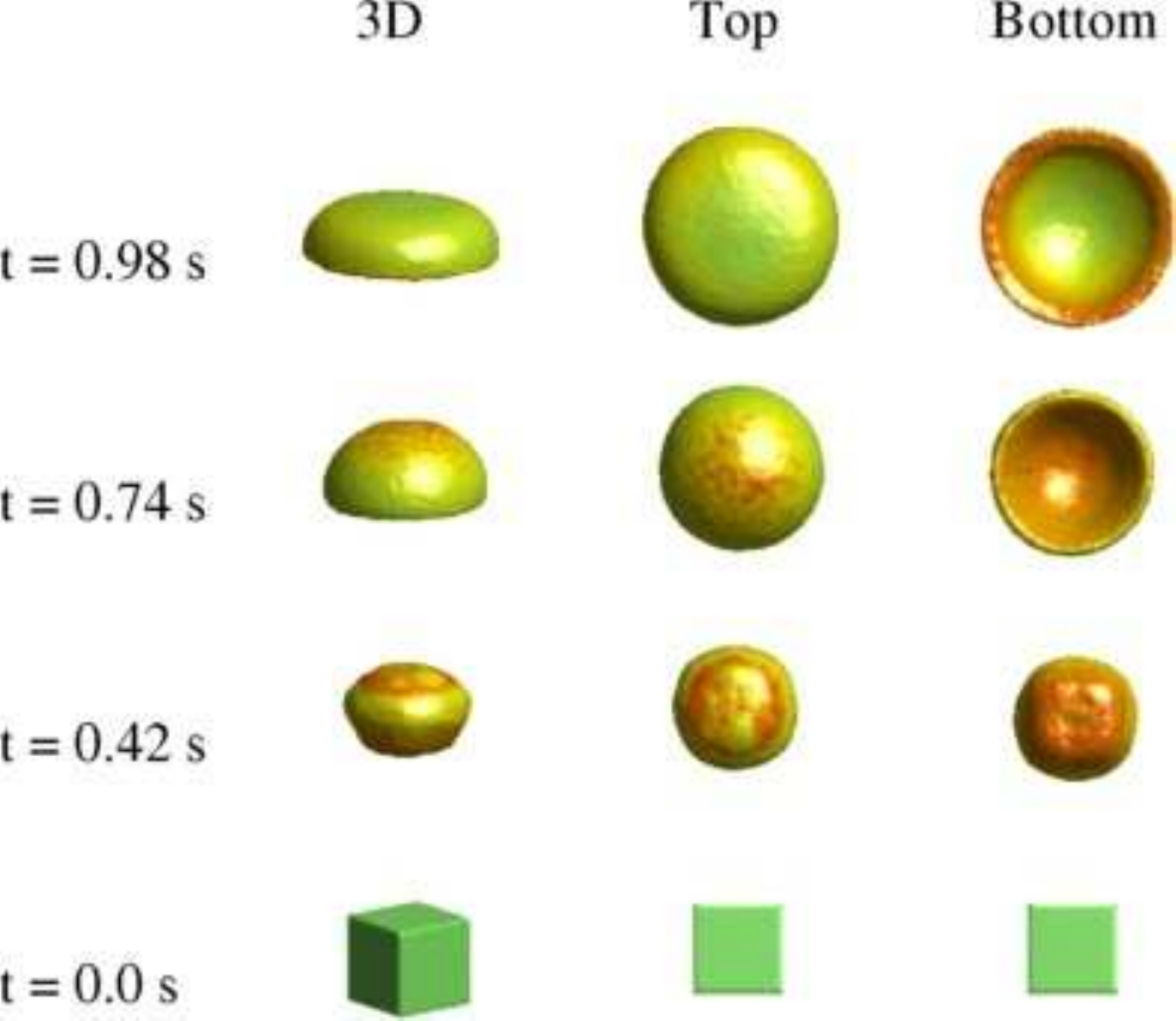}
\caption{GPUSPH simulation of an initially cubic oil drop with side length $L$ = 0.04~m rising through still water with a water depth of 0.2~m. 3D view (left column); top view (middle column); bottom view (right column).}
\label{fig:oil_cube}
\end{figure}

\begin{figure*}
\centering
\includegraphics[width=0.85\textwidth]{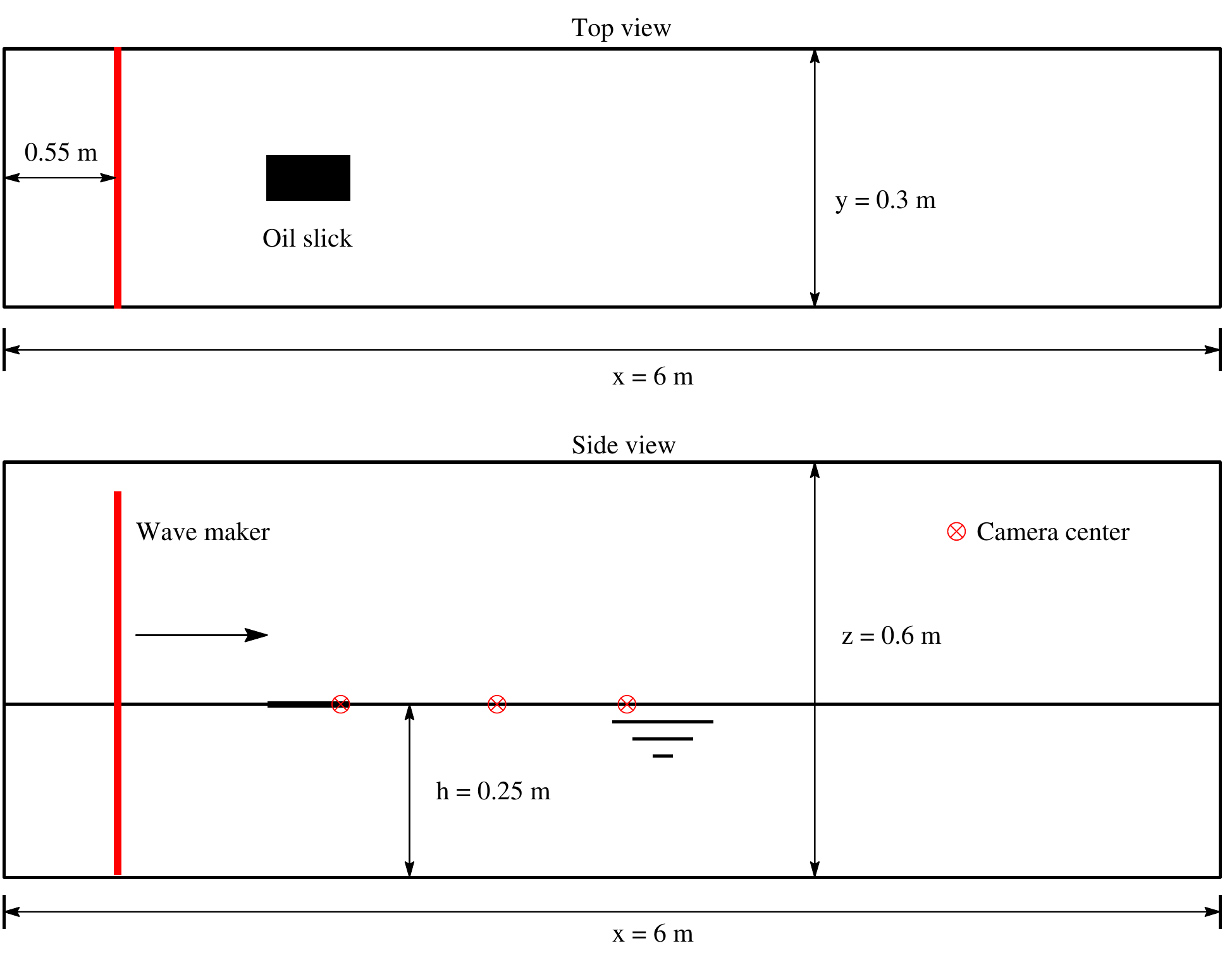}
\caption{Definition sketch of the laboratory experiment on oil dispersion under breaking waves conducted by Li \& Katz~\cite{li2016} at Johns Hopkins University. The cameras are numbered as 1, 2, and 3 from left to right.}
\label{fig:lab_sketch}
\end{figure*}

Prior to applying the two-phase SPH model to simulate a laboratory experiment on oil dispersion under breaking waves, two numerical experiments are carried out to examine the capability of the model for simulating the interaction between immiscible fluids (e.g., oil and water) that are subjected to interfacial surface tension effect. Both cases consider an oil drop rising through a still water tank with a side length equal to $l$ = 0.2~m, and a water depth of $h$ = 0.2~m. The shape of the first oil drop is a sphere with a diameter of $d$ = 0.04~m; the shape of the other oil drop is a cube with a side length of $L$ = 0.04~m. The oil drops are placed near the bottom of the tank with their vertical centers at $z$ = -0.16~m.

An artificial (or numerical) oil-water interfacial surface tension coefficient $\alpha_s$ = 0.05~N/m is used. A particle size of $d_p$ = 0.005~m is used to discretize the water tank and the oil drops. Regarding the numerical boundary condition, the dynamic boundary condition of Dalrymple and Knio~\cite{dalrymple2000sph} is used for the water tank boundary. This boundary condition establishes several rows of boundary particles, and these dynamic boundary particles share the same equations of continuity and state as the fluid particles placed inside the domain; however, their positions and velocities remain unchanged in time. It should be noted that this study includes two types of fluids (i.e., water and oil), but only the property of water is assigned to the dynamic boundary particles, since our simulation does not involve the interaction between oil and the water tank wall.

Fig.~\ref{fig:oil_sphere} shows the shape change of the initially spherical oil drop rising through the water column. Multiple time levels and three view angles are provided. Once the oil drop is released, it starts to rise toward the water surface, and it adjusts its shape due to the buoyancy force and the oil-water interfacial surface tension effect. At $t$ = 0.42~s, the initial sphere was prolonged in the vertical direction, resulting in an ellipsoid. Then the oil ellipsoid turns into a mushroom-shaped oil cloud with a thin tail, as seen at $t$ = 0.74~s. Gradually, the head of the oil cloud becomes bigger and bigger, and eventually the tail is swallowed into the head, and the oil cloud turns into a smooth cone, as indicated at $t$ = 1.04~s. Eventually, the oil cloud reaches the water surface, it becomes a flat-pan oil slick at $t$ = 1.22~s. Fig.~\ref{fig:oil_cube} shows the shape change of the initially cubic oil drop rising through the water column under the effect of the buoyancy force and the oil-water interfacial surface tension effect. Although there is a slight difference regarding the initial shape change between the oil cube and the oil sphere, eventually the initially cubic oil drop also becomes a similar flat-pan oil slick, as observed in the initially spherical oil drop case.

It should be pointed out that although the two numerical experiments show that GPUSPH is able to predict realistic shape changes of oil drops rising through still water under the effect of buoyancy force, our future effort will be made to validate the capability of the model by comparing with laboratory measurements or analytical solutions.
\begin{figure}
\centering
\includegraphics[width=2.75in]{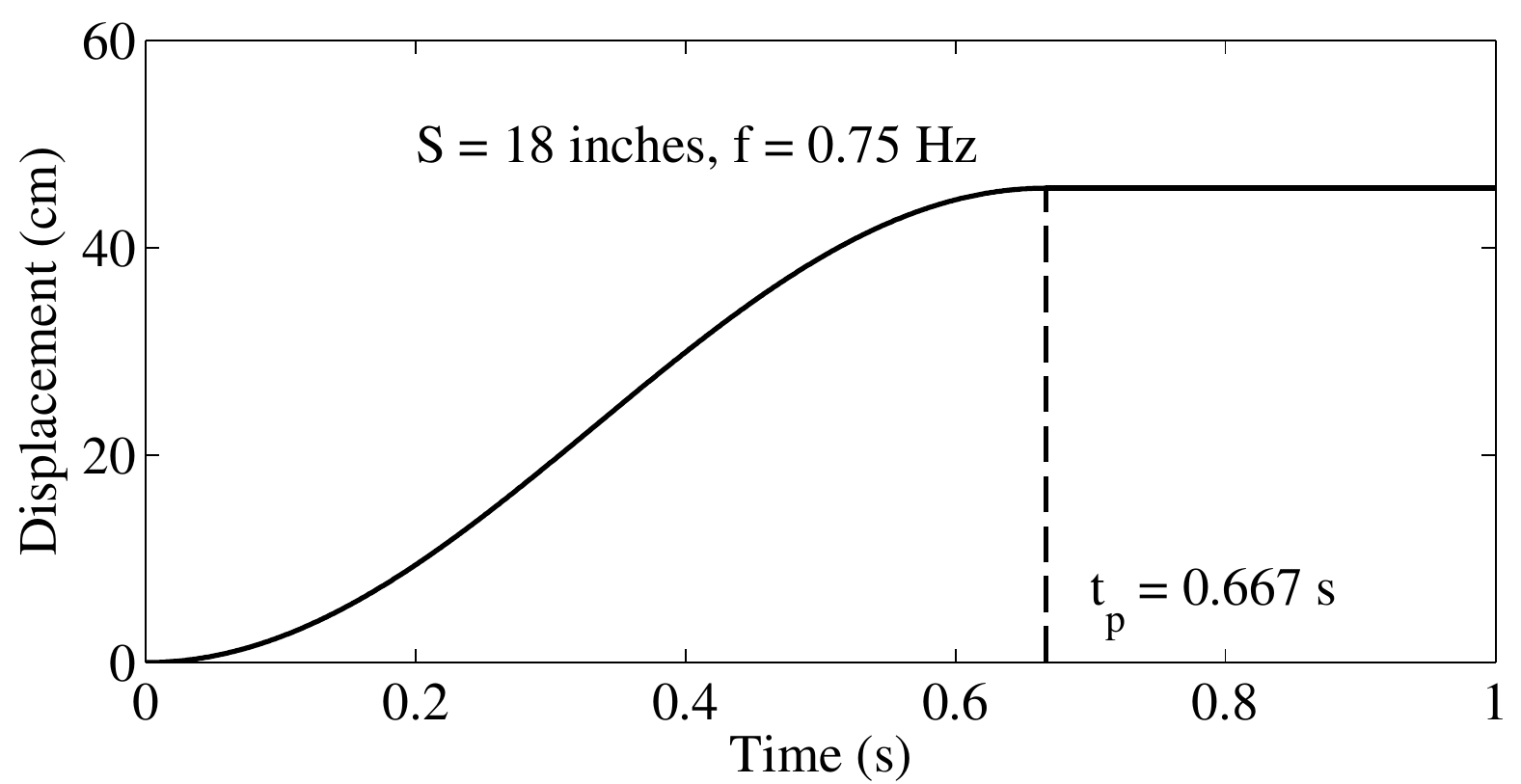}
\caption{The piston-type wave maker trajectory.}
\label{fig:paddle_motion}
\end{figure}

\begin{figure*}
\centering
\includegraphics[width=\textwidth]{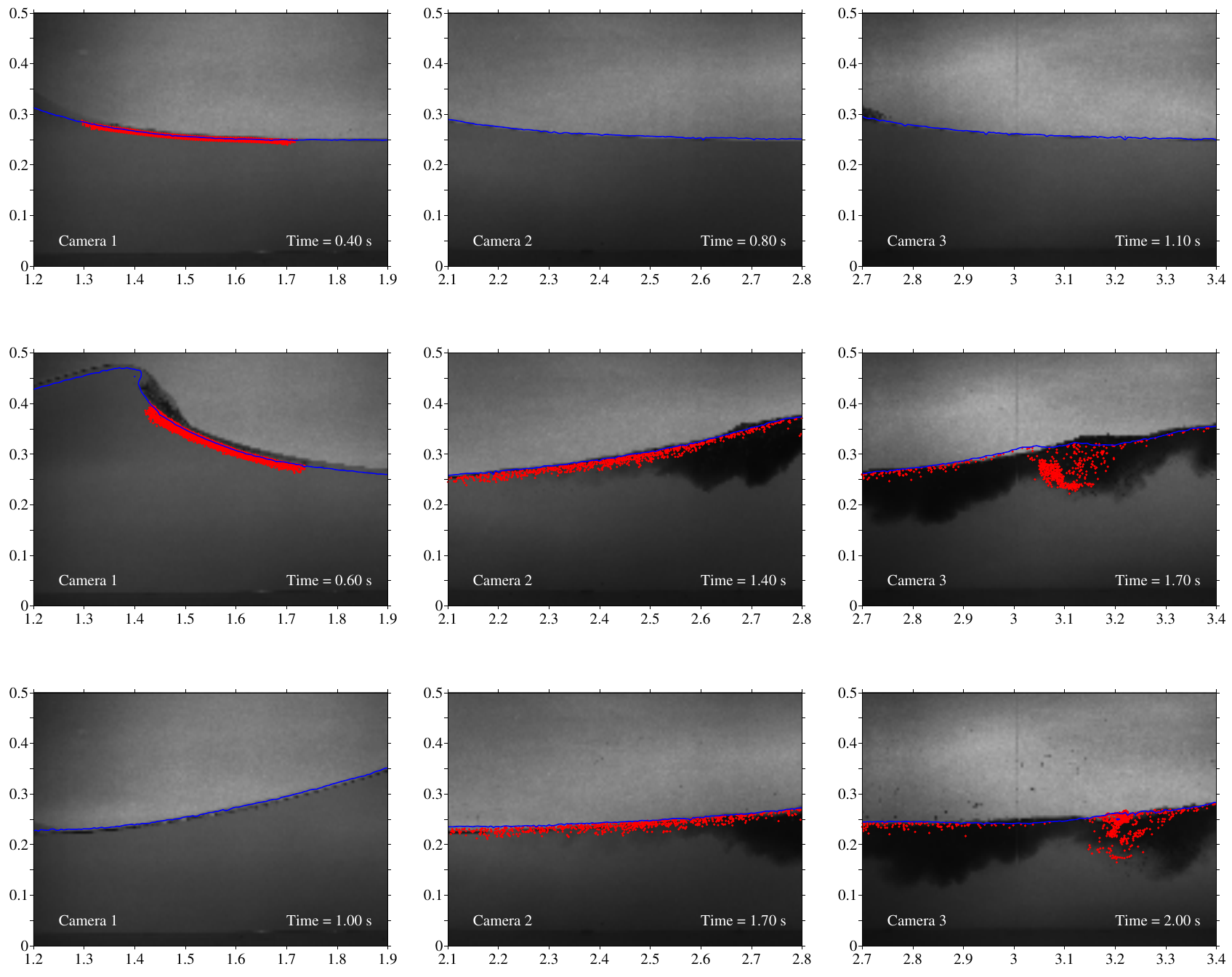}
\caption{Comparison of numerical predictions on oil dispersion under non-breaking wave with laboratory images. Oil is in black in the laboratory; blue line is free surface in GPUSPH; and red dots are oil particles in GPUSPH (for visualization, numerical oil particle in the plot is not in true scale).}
\label{fig:waveform}
\end{figure*}

\section{Laboratory experiment \& numerical setup}\label{lab_model}

A series of laboratory experiments have been conducted in the Department of Mechanical Engineering of Johns Hopkins University by Li \& Katz~\cite{li2016} to investigate oil dispersion under breaking waves. Fig.~\ref{fig:lab_sketch} shows the laboratory experimental setup. The wave flume is about 6~m long, 0.3~wide, and 0.6~m high. This study considers a case with a water depth of $h$ = 0.25~m, and a piston-type wave maker is initially placed around $x$ = 0.55~m from the beginning of the tank. The motion of the wave maker is shown in Fig.~\ref{fig:paddle_motion}, and it is determined by the following equation:
\begin{equation}
x(t) = \frac{45.72}{2}-\frac{45.72}{2}\cos (2\pi\cdot 0.75t)~\text{cm} \quad 0\leq t\leq \frac{2}{3}~\text{s}
\end{equation}

Basically, the driven signal is a $cosine$ function with a frequency equal to 0.75~Hz, and the total displacement of the wave maker is 18~inches (45.72~cm). A 10 $\times$ 1 inch (25.4 $\times$ 2.54~cm) oil slick was placed in the water surface by an oil confinement system just before the touch-down of the breaking wave. The center of the oil slick is about $x$ = 1.5~m and $y$ = 0.15~m, and the thickness of the oil slick is only about 0.5~mm. The measured interfacial surface tension coefficient of the crude oil is about 0.019~N/m. In the laboratory experiments, the oil dispersion under a breaking wave was recorded by three cameras, which are denoted as 1, 2, and 3 from left to right in this study. In the following section, the numerical prediction of the two-phase model is compared with their laboratory images.

In GPUSPH, the 6~m long tank length was fully reproduced, but the tank width (0.3 m) was narrowed and a cross-tank periodic boundary condition was used to save computation cost. Two particle sizes of 1~mm and 5~mm were used to discretize the wave tank, and there are four types of particles: the tank wall, the piston wavemaker, fluid (water), and fluid (oil). For the numerical wall boundary condition, the dynamic boundary condition of Dalrymple and Knio \cite{dalrymple2000sph} was also used. The numerical simulations were carried out by running GPUSPH on 6 NVIDIA Tesla C2050 GPUs at the Graphics Processing Lab Cluster of Johns Hopkins University. Two seconds physical time was considered; the total number of particles ranges from 0.5 million to 20 million, and it took about several hours/several days to finish the simulations.

\section{Results \& Discussions}\label{results_discussions}
Fig.~\ref{fig:waveform} shows the comparison of non-breaking wave form with the laboratory images for all three cameras. Camera~1 is the nearest one to the wave maker, and Camera~3 is the furthest one. For Camera~1, the first two frames ($t$ = 0.4 and 0.6~s) show the comparison of the incoming wave before breaking, and the last frame ($t$ = 1.0~s) shows the comparison about the lee side of the broken wave. For Camera~2, the first frame ($t$ = 0.8~s) shows comparison of the arrival of wave, and the last two frames ($t$ = 1.4 and 1.7~s) show comparison of the passing of the broken wave. Similarly, the first frame ($t$ = 1.1~s) of Camera~3 shows comparison of the arrival of wave, and the last two frames ($t$ = 1.7 and 2.0~s) show comparison of the passing of the broken wave. It is seen that GPUSPH accurately predicts the pre-breaking and post-breaking wave profiles as recorded by all three cameras in the laboratory. 

\begin{figure}
\captionsetup[subfigure]{labelformat=empty}
\begin{minipage}{.5\linewidth}
\centering
\subfloat[]
{\includegraphics[width=3.15in]{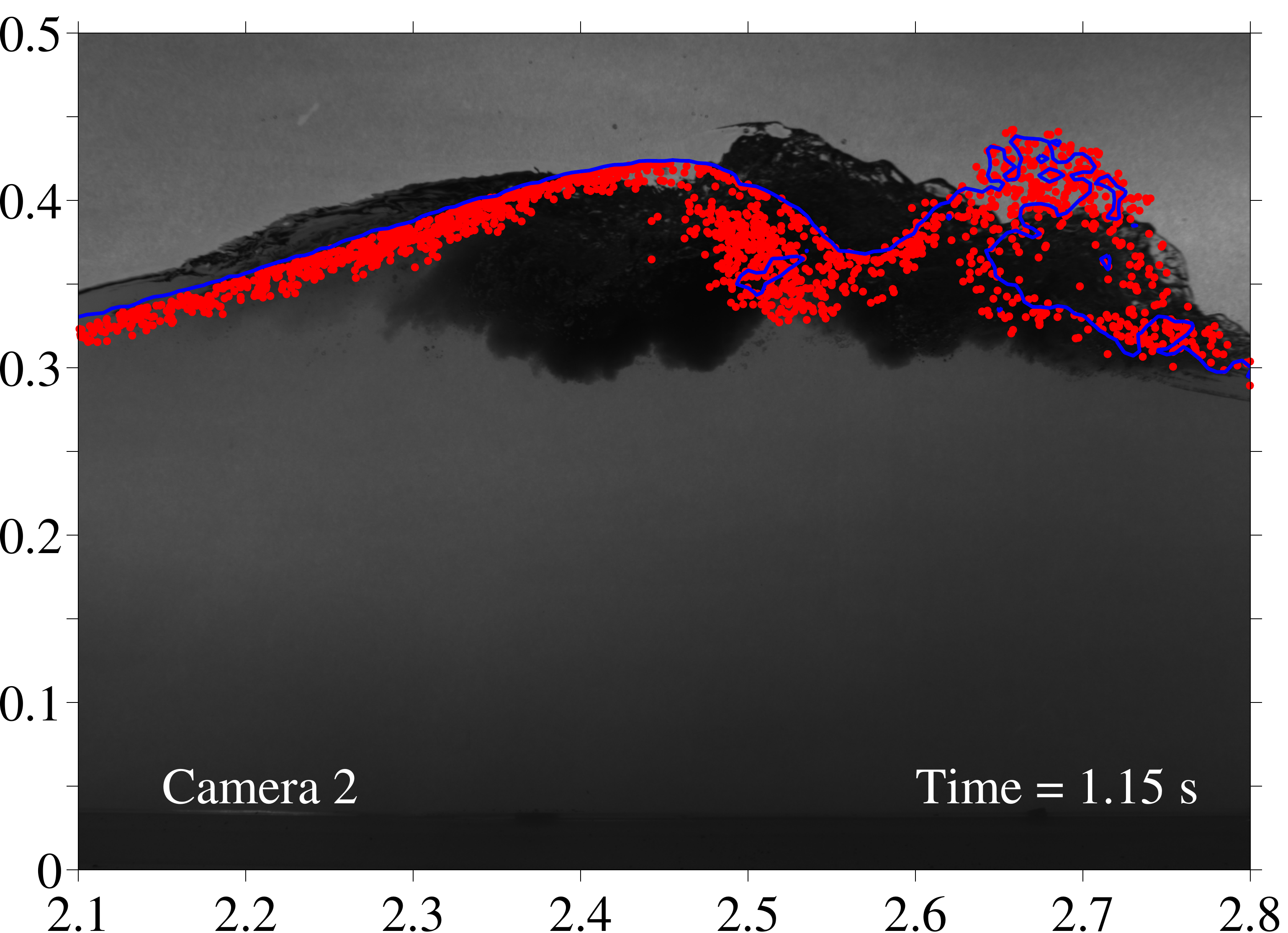}}
\end{minipage}
\begin{minipage}{.5\linewidth}
\centering
\subfloat[]
{\includegraphics[width=3.15in]{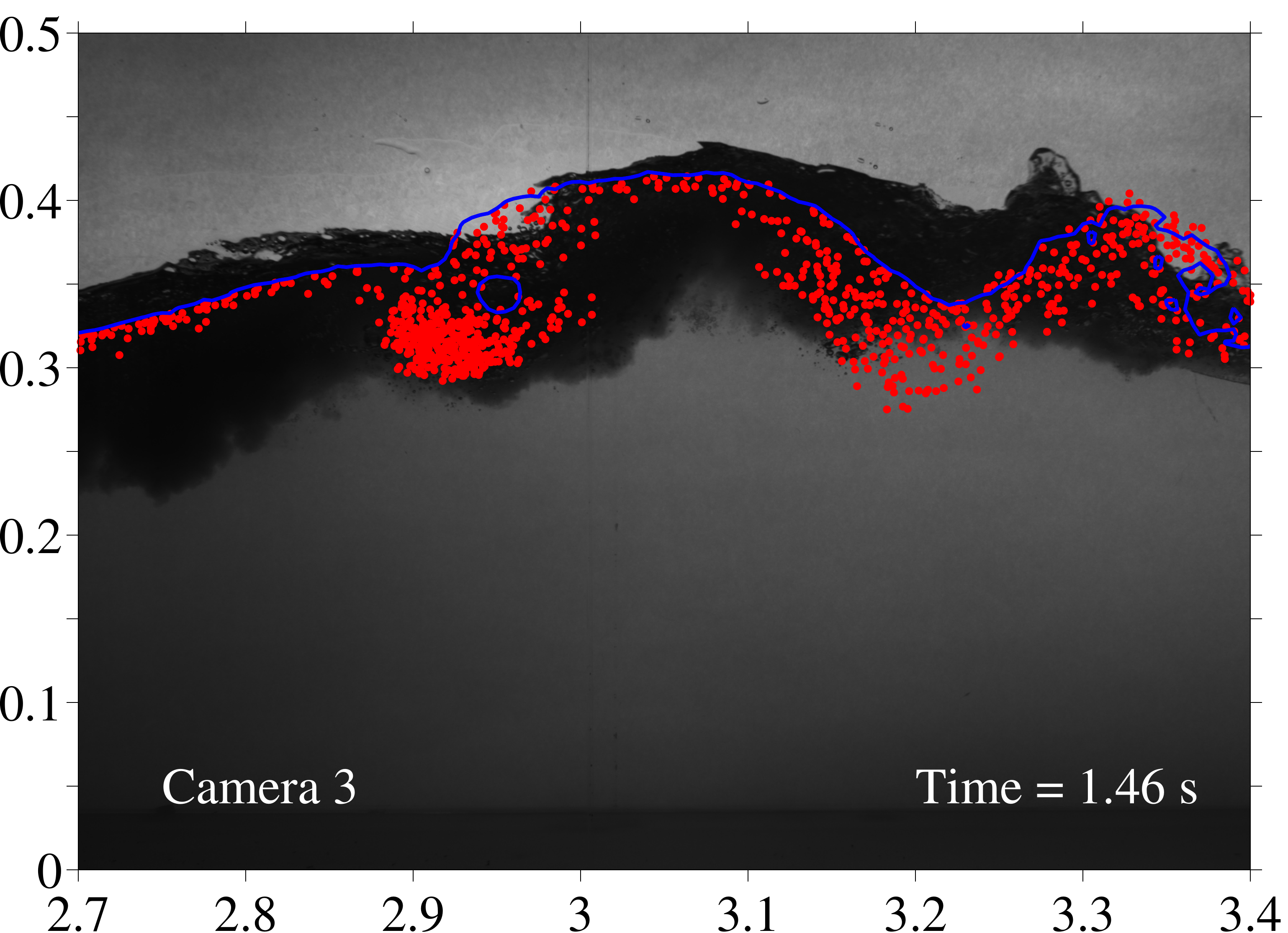}}
\end{minipage}
\caption{Comparison of numerical predictions on oil dispersion under breaking wave with laboratory images. Oil is in black in the laboratory; blue line is free surface in GPUSPH; and red dots are oil particles in GPUSPH (for visualization, numerical oil particle in the plot is not in true scale).}
\label{fig:wavebreaking}
\end{figure}

However, there are also discrepancies observed between the numerical prediction and the laboratory record. For the oil distribution, although good agreements are obtained before the wave breaks (e.g., the first two frames of Camera~1 as shown in Fig.~\ref{fig:waveform}), there is a clear mismatch of the submerged oil dispersion after the wave breaks (e.g., the last two frames in both Cameras~2 and 3 as shown in Fig.~\ref{fig:waveform}). Fig.~\ref{fig:wavebreaking} shows the comparison of the breaking wave profiles and the corresponding oil distribution in Cameras~2 and 3. It is seen that the breaking portion of the wave in the laboratory is not the same as produced by GPUSPH nor is the distribution of oil within the water column.

\begin{figure}
\centering
\includegraphics[width=2.75in]{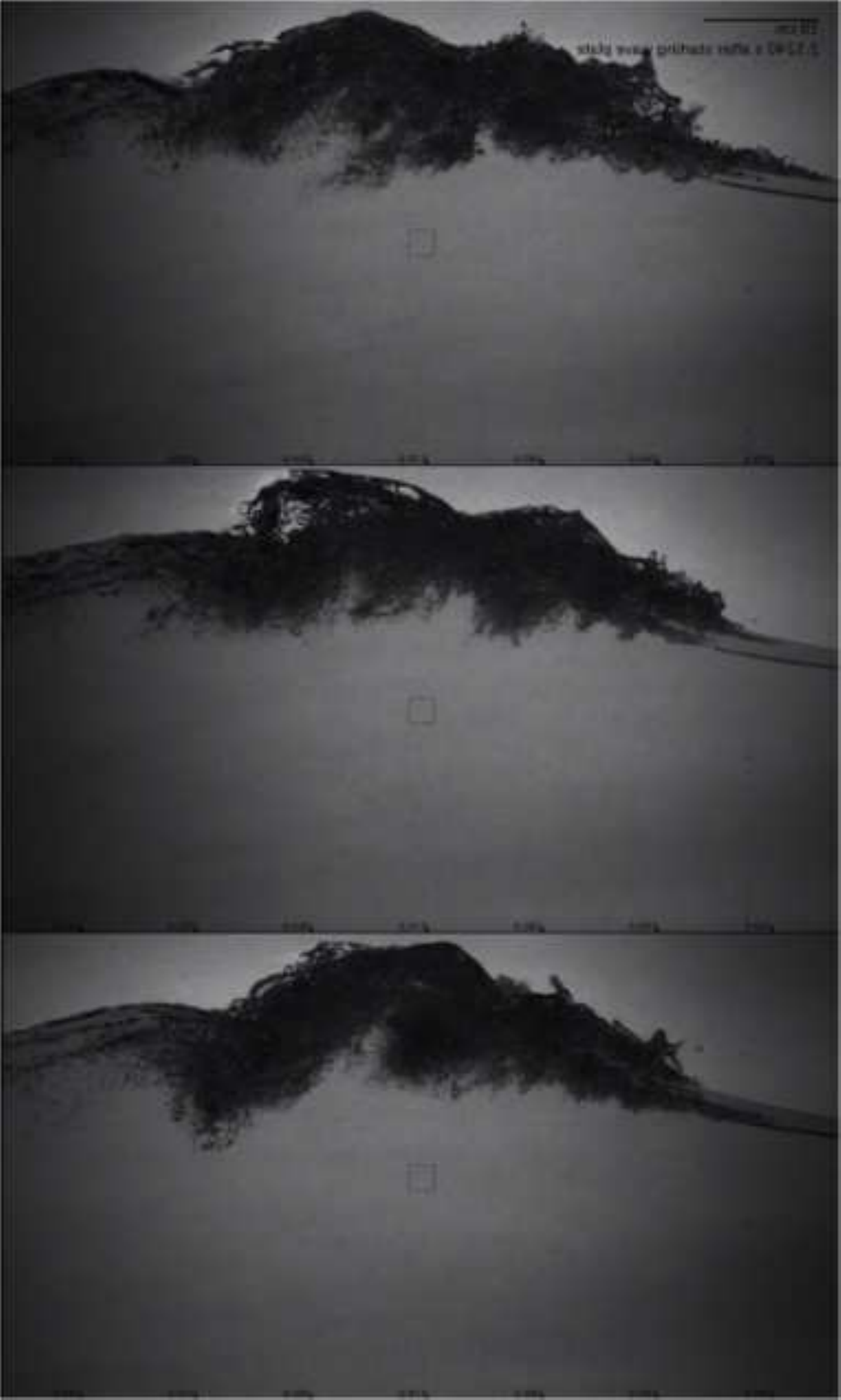}
\caption{Comparison of recorded breaking waves and oil dispersion at the same instant among 3 runs with the same initial laboratory setup. The wave propagates from left to right, the oil is in black in the laboratory experiments. (Image courtesy: Cheng Li).}
\label{fig:lab_wave}
\end{figure}

There are several possible causes for the mismatch of the simulated oil distribution with that recorded in the laboratory. One is related to the numerical setup. Due to the relatively thin oil slick in the laboratory, the numerical model has to use a fine particle size; however, a particle size such as 1~mm is still much larger than the oil droplet size after the oil break-up. Furthermore, the wave breaking is a 3D process in the laboratory, but the high computation cost caused by the fine particle size in the numerical simulation prevents the model to consider the full width of the tank, therefore, the breaking wave profile might not be correctly captured by the numerical model. Last but not the least, the oil distribution is significantly influenced by the breaking wave, but the breaking wave is a chaotic process in nature. Fig.~\ref{fig:lab_wave} shows the comparison of recorded breaking waves and oil dispersion at the same instant among 3 runs with the same initial laboratory setup, it is seen that the breaking wave and oil dispersion could not even be reproduced in the laboratory. Similarly it is not likely that our numerical prediction (even with a fully resolved oil slick thickness by using a very fine particle size in the simulation) could match the breaking wave and the recorded oil dispersion perfectly. Therefore, we can only expect a qualitatively similar result at best.

\section{Conclusions}\label{conclusions}
The study has developed a two-phase SPH model based on a single-phase SPH model GPUSPH, and further applied it to investigate oil dispersion under breaking waves. Major work and findings include:
  \begin{itemize}
  \item
  The two-phase SPH model solves one set of governing equations for both phases. Particles only contribute their volumes in the continuity equation, and renormalization is conducted to preserve density near a free surface. A numerical surface tension model was implemented to take into account the interfacial surface tension effect between two phases;
  \item 
Two numerical examples were carried out to evaluate the model's capability of modeling oil-water interaction. Numerical predictions show realistic shape changes as oil drops rising through still water. However, further model validations with laboratory experiments are needed;
  \item
    The two-phase SPH model was then applied to replicate a laboratory experiment on oil dispersion under breaking waves. It is seen that it is able to reproduce well the pre- \& post-breaking wave as recorded in the laboratory;
  \item
    Oil dispersion predicted by the two-phase SPH model only matches part of the laboratory observation. Several factors (e.g., 3D \& chaotic nature of breaking waves, numerical setup) cause the discrepancy.
  \end{itemize}

\section*{Acknowledgment}
This research was made possible in part by a grant from The Gulf of Mexico Research Initiative, and in part by a grant from Office of Naval Research. Data are publicly available through the Gulf of Mexico Research Initiative Information \& Data Cooperative (GRIIDC) at https://data.gulfresearchinitiative.org.




%



\end{document}